%% file: PPMCISU.tex
\pdfoutput=1
\documentclass[%
preprint,
 amsmath,amssymb,
 aps,
]{revtex4-2}

\usepackage{graphicx}
\usepackage{dcolumn}
\usepackage{bm}

\newcommand\Ref[1] {Ref.\,\cite{#1}}

\newcommand\eqn[1] {Eq.\,(\ref{#1})}

\newcommand\fig[1] {Fig.\,{\ref{#1}}}

\newcommand\sectioning[1]{\section{#1}}

\def\beq{\begin{equation}}
\def\eeq{\end{equation}}
\def\bsp#1\esp{\begin{split}#1\end{split}}
\def\bal#1\eal{\begin{align}#1\end{align}}
\newcommand\lp   {\ensuremath{\left}}
\newcommand\rp   {\ensuremath{\right}}

\newcommand\rO   {\ensuremath{\mathrm{O}}}

\newcommand\GeV  {\ensuremath{\mathrm{GeV}}}

\newcommand\ri   {\ensuremath{\mathrm{i}}}
\newcommand\rt   {\ensuremath{\mathrm{t}}}
\newcommand\mt   {\ensuremath{m_{\rt}}}
\newcommand\muinf{\ensuremath{\mu_{\mathrm{inf}}}}
\newcommand\detL {\ensuremath{{\det}_{C}}}
\newcommand\lSM {\ensuremath{{\lambda}_{\rm SM}}}

\begin{document}

\preprint{
}

\title{Particle physics model of curvaton inflation\\in a stable universe}

\author{Zolt\'an P\'eli}%
 \email{zoltanpeli92@gmail.com}
\author{Istv\'an N\'andori}%
 \email{nandori.istvan@science.unideb.hu}
\affiliation{%
 MTA-DE Particle Physics Research Group,\\ H-4010 Debrecen, PO Box 105, Hungary 
}%

\author{Zolt\'an Tr\'ocs\'anyi}
 \email{Zoltan.Trocsanyi@cern.ch}
 \homepage{http://pppheno.elte.hu/}
\affiliation{
 Institute for Theoretical Physics, ELTE E\"otv\"os Lor\'and University,
P\'azm\'any P\'eter s\'et\'any 1/A, H-1117 Budapest \\
and MTA-DE Particle Physics Research Group
}%

\date{\today}

\begin{abstract}
We investigate a particle physics model for cosmic inflation based on the following assumptions: (i) there are at least two complex scalar fields; (ii) the scalar potential is bounded from below and remains perturbative up to the Planck scale; (iii) we assume slow-roll inflation with maximally correlated adiabatic and entropy fluctuations 50--60 e-folds before the end of inflation. The energy scale of the inflation is set automatically by the model. Assuming also at least one massive right handed neutrino, we explore the allowed parameter space of the scalar potential as a function of the Yukawa coupling of this neutrino.
\end{abstract}

\maketitle


\sectioning{Introduction}
\label{sec:intro}
\input{intro}

\sectioning{Particle physics model}
\label{sec:model}
\input{model}

\sectioning{Cosmological inflation}
\label{sec:cosmology}
\input{cosmology}

\sectioning{Predicting the scalar couplings}
\label{sec:predictions}
\input{parameterspace}

\sectioning{Conclusions}
In this letter we proposed a particle physics model of cosmic inflation. It requires at least 
two scalar fields. We found that in a small region of the parameter space of the scalar couplings, the determinant 
of the scalar quartic coupling matrix becomes very small at a scale around $10^{16}$\,GeV. 
As a result the global minimum of the scalar potential increases significantly, allowing for an accelerated 
expansion of the universe by a slow-roll model at this scale, called the scale of inflation. 
We assume the curvaton scenario of inflation, i.e.~maximally correlated adiabatic and entropy fluctuations at 
$50-60$ e-folds before the end of inflation, which implies vanishing tensor-to-scalar ratio. To set the 
normalization of the potential at vanishing field values, we required that the model reproduces the
measured value of the scalar tilt. The inflation stops when the parameter that measures
the acceleration of the fields starts to increase quickly.
After this the global minimum of the potential decreases preventing the appearance of another 
period of inflation.

\begin{acknowledgments}
This work was
supported by grant K 125105 of the National Research, Development and
Innovation Fund in Hungary.
\end{acknowledgments}


\nocite{*}

\bibliography{PPMCISU}

\end{document}

%% file: intro.tex
The precise mechanism of cosmic inflation is one of the most pressing
questions in our understanding of the early universe. Today the original 
idea for inflation \cite{Guth:1980zm} is not favoured because it is unclear 
how to define a proper mechanism to explain the required reheating of the
universe. A popular solution to this question of reheating is the slow-roll
scenario \cite{Linde:1981mu,Albrecht:1982wi} in which the ground state starts
from an unstable position and rolls down very slowly to a local or global minimum.
The inflation stops when the potential energy function becomes too steep, which 
leads to a fast roll. 
In principle, the slow roll can start from a large field value and proceed 
towards a minimum with a smaller field value, or from a small (essentially vanishing)
field value to a larger minimum. These two cases are referred to as 
large- and small-field slow roll \cite{Baumann:2009ds}.
A problem related to large-field slow-roll is the initial value problem, 
namely one has to explain why the ground state starts from a value much 
larger than the typical energy scale of inflation. Chaotic inflation
\cite{Linde:1986fd} was devised to handle this problem, but then one 
has to assume very large -- again larger than the scale of inflation 
-- fluctuations. 
The origin of inflation is still an open question in
cosmology \cite{Earman:1999zia,Steinhardt:2011zza}.  

It is known that scalar fields can mimic the equation of state required 
for the exponential expansion of the early universe \cite{Linde:1981mu,Albrecht:1982wi}. As the
Higgs boson was discovered \cite{Aad:2012tfa,Chatrchyan:2012xdj}, we know that at least one doublet scalar field
exists in nature. Hence, it may appear natural to assume that the
Brout-Englert-Higgs (BEH) field is the inflaton (see for example \Ref{Bezrukov:2014ipa}), but such a 
scenario was criticised, see for instance \Ref{Martin:2013tda}. Many types of scalar potentials 
have already been discussed in the literature as viable scenarios 
for cosmic inflation \cite{Martin:2013tda}.  There are three major categories 
of scalar inflaton potentials with minimal kinetic terms: (i) the large 
field, (ii) the small field and (iii) the hybrid models. In the third 
case one introduces more than one field, with one of those being
the inflaton and the other field switches off the exponential expansion. In this
sense it is not a real multifield model. The case of hybrid models
is excluded by experimental observations because those predict a 
scalar tilt $n_s$ larger than one in contradiction with
the observed structure of the thermal fluctuations of the cosmic 
microwave background radiation (CMBR) resulting in $n_s = 0.9677\pm 0.0060$  
\cite{Ade:2015lrj,Array:2015xqh}. The tensor and scalar power spectra of the CMBR
suggest a small value for the tensor-to-scalar ratio $r$, consistent with zero,
which emerges automatically in real multifield models with
curvaton scenario \cite{Mollerach:1989hu,Linde:1996gt,Lyth:2001nq,Kitajima:2017fiy}.

In this letter we consider a simple multifield particle physics model
of cosmic inflation with a curvaton scenario. We show that in a fairly constrained region of the parameter space, the model can provide a natural switch on and off mechanism of inflation. 

%% file: model.tex
The particle content of the model coincides with that in the standard model of particle interactions, supplemented with one complex scalar field. We also allow for one (or more) Dirac-, or Majorana-type right-handed neutrinos.
In this letter for the sake of definiteness we consider the case of Dirac neutrinos.

In addition to the usual $SU(2)$-doublet scalar field
\begin{equation}
\phi=\lp(\!\!\begin{array}{c}
               \phi^{+} \\
               \phi^{0}
             \end{array}\!\!\rp) = 
\frac{1}{\sqrt{2}}
\lp(\!\!\begin{array}{c}
          \phi_{1}+\ri\phi_{2} \\
          \phi_{3}+\ri\phi_{4}
        \end{array}\!\!
\rp)
\,,
\end{equation}
we assume the existence of a complex scalar $\chi$ that transforms as a
singlet under the standard model gauge transformations. The potential 
energy of these scalar fields is assumed as
\beq
V(\phi,\chi) = V_0 - \mu_\phi^2 |\phi|^2 - \mu_\chi^2 |\chi|^2
+ \frac12 \lp(|\phi|^2, |\chi|^2\rp)
C
\lp(\!\!\begin{array}{c}
|\phi|^2 \\ |\chi|^2
\end{array}\!\!\rp)
\label{eq:V}
\eeq
where $|\phi|^2 = |\phi^+|^2 + |\phi^0|^2$ and
$
C =
\lp(\!\!\begin{array}{cc}
 2\lambda_\phi & \lambda \\ \lambda & 2\lambda_\chi
\end{array}\!\!\rp)
$
is the coupling matrix. This potential energy function
contains a coupling term $\lambda |\phi|^2 |\chi|^2$ of the scalar fields 
in addition to the usual quartic terms.
The value of the additive constant $V_0$ is irrelevant for particle
dynamics, but as we shall see, it is relevant for the inflationary 
model, hence we allow a non-vanishing value for it. 
In order that this potential energy be bounded from below, we have to 
require the positivity of the self-couplings, $\lambda_\phi$, 
$\lambda_\chi>0$, and also the coupling matrix be positive definite,
\beq
\detL = 4\lambda_\phi \lambda_\chi - \lambda^2 > 0
\,,\textrm{  if   } \lambda < 0
\,.
\label{eq:positivity}
\eeq
Our model for cosmic inflation works only if $\lambda<0$%
\footnote{For $\lambda>0$ the global minimum of $V$ falls on one of the coordinate axes when $\detL$ becomes small. As a result our inflationary scenario does not work.}.
If these conditions are satisfied, we find the minimum of the
potential energy at field values $\phi= v$ and $\chi = w$ where 
the vacuum expectation values (VEVs) are
\beq
v = \sqrt{\frac{2\lambda_\chi \mu_\phi^2 - \lambda \mu_\chi^2}
{4 \lambda_\phi \lambda_\chi - \lambda^2}}
\,,\qquad
w = \sqrt{\frac{2\lambda_\phi \mu_\chi^2 - \lambda \mu_\phi^2}
{4 \lambda_\phi \lambda_\chi - \lambda^2}}
\,.
\label{eq:VEVs}
\eeq
Using the VEVs, we can express the quadratic couplings as
\beq
\mu_\phi^2 = 2 \lambda_\phi v^2 + \lambda w^2
\,,\qquad
\mu_\chi^2 = 2 \lambda_\chi w^2 + \lambda v^2
\,.
\label{eq:scalarmasses}
\eeq
For $\lambda < 0$, the
constraint (\ref{eq:positivity}) ensures that the denominators of the
VEVs in \eqn{eq:VEVs} are positive, so the VEVs have non-vanishing real
values only if the inequalities
\beq
2\lambda_\chi \mu_\phi^2 - \lambda \mu_\chi^2 > 0
\quad\text{and}\quad
2\lambda_\phi \mu_\chi^2 - \lambda \mu_\phi^2 > 0
\label{eq:muXconditions}
\eeq
are satisfied simultaneously, which can be fulfilled  \cite{Peli:2019xwv} if at most one of the quadratic
couplings is smaller than zero
\footnote{In our nodel for inflation we only consider the case when both $\mu^2_i>0$.}.

After spontaneous symmetry breaking and choosing unitary gauge, 
the scalar kinetic term leads to a mass matrix of the two real scalars
\footnote{The would be massless Goldstone boson can give mass to 
a new neutral vector boson as described in \Ref{Trocsanyi:2018bkm}}. 
We can diagonalize this matrix by an orthogonal rotation and find 
for the masses $M_{h/H}$ of the mass eigenstates:
\beq
\frac{M_{h/H}^2}2 = \lambda_\phi v^2 + \lambda_\chi w^2
\mp \sqrt{(\lambda_\phi v^2 - \lambda_\chi w^2)^2 +
(\lambda v w)^2}
\label{eq:MhH}
\eeq
where $M_h \leq M_H$ by convention. At this point either $h$ or $H$ can
correspond to the observed scalar boson. As $M_h$ must be positive, the condition
\beq
v^2 w^2 \Big(4 \lambda_\phi \lambda_\chi - \lambda^2\Big) > 0
\label{eq:Mhpositivity}
\eeq
has to be fulfilled. If both VEVs are greater than zero, as needed for
two non-vanishing scalar masses, then this condition coincides with the
positivity constraint (\ref{eq:positivity}).

We studied the ultraviolet behaviour of the scalar couplings of this model in \Ref{Peli:2019xwv} where we constrained the parameter space by requiring that 
(i) the scalar potential remains bounded from below and
(ii) the couplings remain perturbative up to the Planck scale $m_P$. 

%% file: cosmology.tex
\begin{figure}[h]
\includegraphics[width=\linewidth]{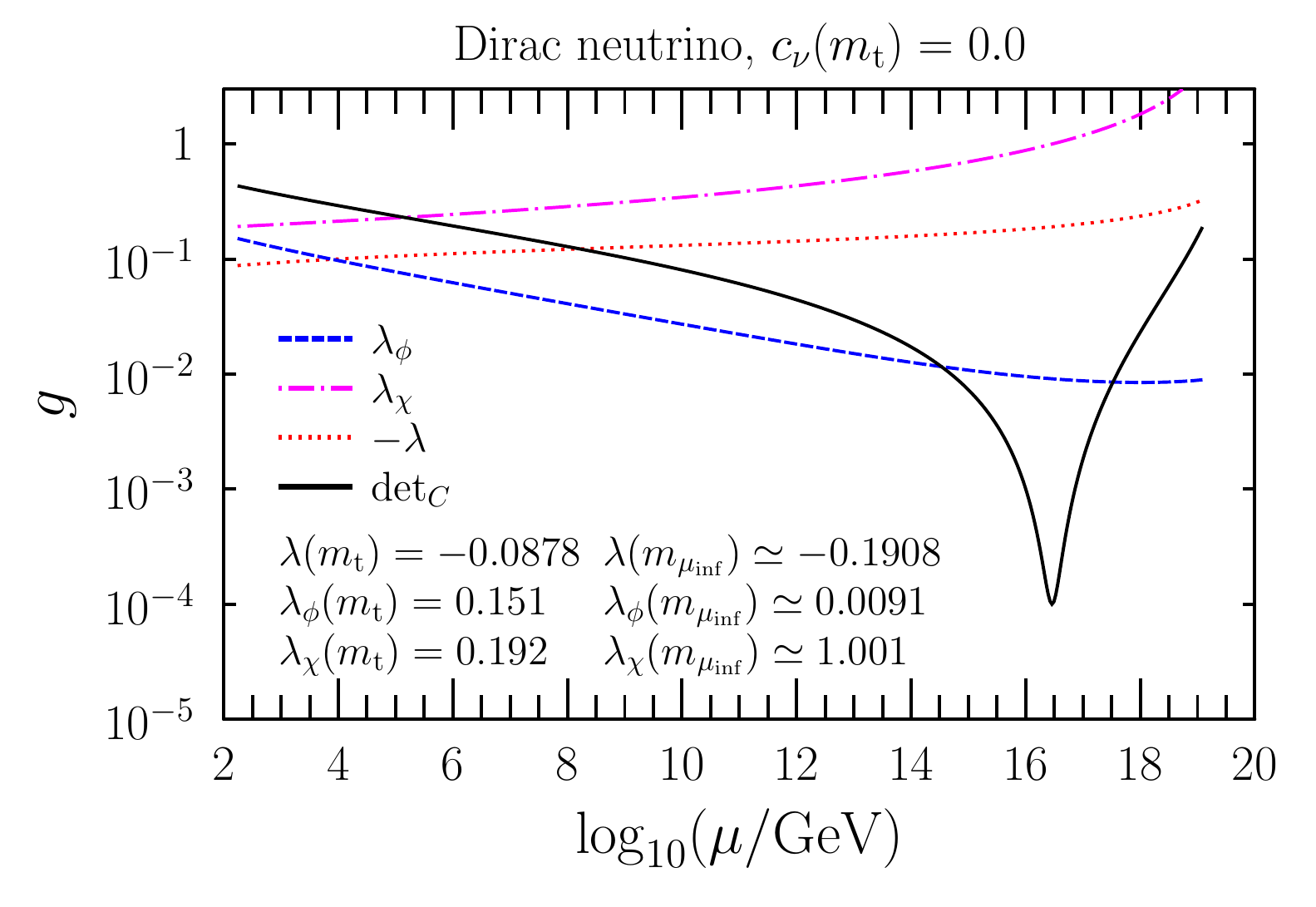}
\caption{\label{fig:running} Running of the scalar couplings and of $\detL$. $g$ means any of the couplings.}
\end{figure}

 We now explore the cosmic inflation of the two-field model with potential energy defined in \eqn{eq:V}.
 We consider slow-roll inflation when the potential energy has a large, almost flat area for small field 
 values and a global minimum at large values of the VEVs. Such a potential energy allows for slow roll 
 of the fields from small values towards the global minimum, resulting in cosmic inflation. 
 The required form of the potential energy function appears naturally at some high energy scale, for certain 
 values of the scalar couplings at the mass of the t-quark $m_\rt$. As \eqn{eq:VEVs} shows, the VEVs are 
 inversely proportional to $\sqrt{\detL}$. 
 \fig{fig:running} shows the running of $\detL$ together with that of the couplings from 
 initial values at $m_\rt$ chosen from the stability region. We see a narrow wedge -- like 
 an inverse resonance -- where $\detL$ becomes very small, implying VEVs at around field values of 
 $10^{5}$\,GeV. The figure shows an example with vanishing Yukawa coupling $c_\nu$ of the right-handed neutrino, 
 but below we show that the value of $c_\nu$ influences only the size of the parameter space of the scalar 
 couplings where this phenomenon leads to such potential energy function that can support cosmic 
 inflation in accordance with current values of relevant observables. 
 
 The single-field inflationary models predict purely curvature perturbations, 
 resulting from energy density fluctuations. Having multiple fields allows for multiple types of fluctuations, 
 hence several observable quantities, such as the tilts corresponding to curvature, 
 isocurvature (emerging due to fluctuations in the relative number density of particles) 
 and 
 a correlation angle $\Delta$ \cite{Byrnes:2006fr}.
 
 Following \Ref{Gordon:2000hv}, we introduce a local rotation of $(\phi,\chi)$ into $(\sigma,s)$ where 
 $\sigma$ refers to the adiabatic field and $s$ to the entropy field. The adiabatic field is the path length 
 along the classical trajectory, while $s$ remains a constant. 
 The number of slow-roll parameters also increase. In the single-field case, in addition to 
 the parameter $\epsilon$ describing the deviation from the equation of state of the de-Sitter space-time,
$\epsilon = \frac{3}{2}\left(\frac{\displaystyle p}{\displaystyle \rho}+1 \right)$, there is only one other slow roll parameter $\eta$, 
which essentially measures the acceleration of the fields. In our example we have three $\eta$ parameters 
that can be expressed approximately from the potential as
\beq
\eta_{ij} \simeq m_p^2 \frac{\partial_{ij}V}{V}\,,
\quad ij = \phi\phi,\:\chi\chi,\:\phi\chi,\:ss,\:\sigma\sigma,\:s\sigma
\label{eq:etaij}
\eeq
(note that $\eta_{\phi\phi}+\eta_{\chi\chi}= \eta_{\sigma\sigma}+\eta_{ss}$), while
\beq
\epsilon \simeq \frac{1}{2}m_P^2 \biggl( \frac{\partial_\sigma V}{V} \biggr)^2
\,.
\eeq
In principle, inflation is possible only until both $\epsilon$ and $\eta_{ij}$ are small, resulting in the slow roll.

\begin{figure}[t]
\includegraphics[width=\linewidth]{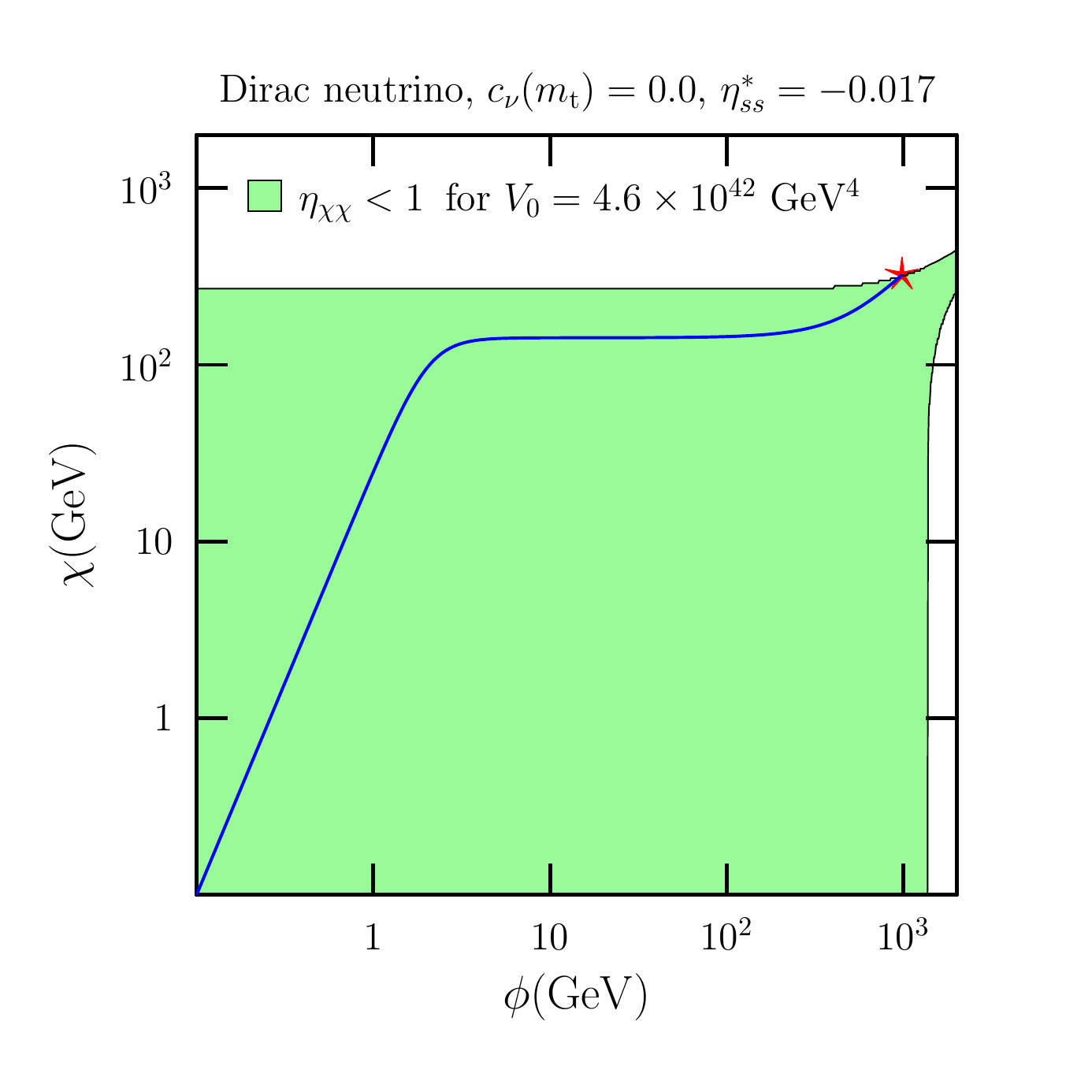}
\caption{\label{fig:trajectory} A possible trajectory of the rolling of the scalar fields.}
\end{figure}

To set the exact conditions of slow roll, we solved the equations of motion with the integration 
variable transformed to the number of e-folds $N$, and terminated the process, when either of the slow-roll 
parameters reached unity. We set the starting point of the trajectory at vanishing field values. For the parameter 
values of the potential energy we used values allowed by the perturbativity and stability conditions mentioned 
in the section describing the model, namely $|\lambda|,\:\lambda_i \sim \rO(10^{-1}-10^{-2})$ and $\mu_i^2/\GeV^2 \sim \rO(1-3 \times 10^{4})$.
For such values we have found that the $\eta_{ij}$ parameters increase much faster than $\epsilon$, 
reaching 1, while $\epsilon$ remaining small, about $\rO(10^{-30})$. Hence, we set the end of inflation by the 
condition $\eta_{ij}=1$. In practice the parameter $\eta_{\chi\chi}$ increases the fastest. We show an example of such a trajectory in \fig{fig:trajectory}. This trajectory induces $N>200$ e-folds. The value of $\eta_{ss}^*$ refers to the value of $\eta_{ss}$ at $60$ e-folds before the end of inflation.

The observables are constructed from the slow-roll parameters taken $50-60$ e-folds before the end of inflation. This corresponds to an even smaller $\epsilon$, which reduces the tensor-to-scalar ratio,
$r = 16 \epsilon (\sin \Delta)^2$
to essentially zero. Such a small $r$ is not excluded by cosmological measurements. The smallness of $r$ however
is in conflict with the traditional cosmological normalization
\begin{equation}
V_0 \simeq r\lp(1.6 \cdot 10^{16} \text{ GeV} \rp)^4
\,.
\end{equation}
This conflict may be resolved by assuming that the adiabatic and entropy fluctuations were maximally correlated at $50-60$ 
e-folds before the end of inflation, implying $\cos\Delta=1$, and hence predicting zero for the tensor-to-scalar ratio. 
Consequently, we have to find different conditions to set the scale of inflation $\muinf$ and for 
the normalization of the potential energy. As suggested above, we provide the first from the particle physics 
model by identifying $\muinf$ with the location of the wedge in the running of $\detL$. 
The case of $\Delta=0$, i.e.~maximally correlated fluctuations are referred to as the curvaton scenario. In this case, the various 
tilts coincide. Neglecting $\epsilon$, we have:
\begin{equation}
n_s -1 = 2 \eta_{ss}
\end{equation}
Considering $\eta_{ss}$ as a function of $V_0$ (see \eqn{eq:etaij} with $V_0$ in $V$ in the denominator), we normalize it to produce the scalar tilt in agreement 
with the most recent data, $n_s\simeq 0.966$, yielding $V_0 \simeq 5 \times 10^{42}\,\GeV^4$.

Having fixed the value of $V_0$, we propose the following inflationary scenario. The scalar potential energy is
given by \eqn{eq:V}. After the Big Bang the characteristic energy scale of particle interactions is 
near the Planck scale, hence the scalar fields are fluctuating around zero.
As the universe expands, the characteristic energy scale decreases and the scalar couplings run according
to their renormalization group equations, exhibiting the wedge for $\detL$ at a scale $\muinf$ (around 
$10^{16}\,$GeV) that we identify with the scale of inflation. At this scale the global minimum of the
potential energy function increases to about $10^5$\,GeV and the fields start to roll slowly towards this minimum,
resulting in cosmic inflation. This accelerated expansion continues until the acceleration (second time derivative of the fields) remain negligible in the equation of the motion, determined by $\max \eta_{ij}=1$.
The universe starts its Hubble expansion, decreasing the characteristic energy scale, and the global minimum 
of the scalar potential quickly returns to small field values.

%% file: parameterspace.tex
The cosmological inflation as described in the previous section occurs only in a restricted region of the parameter space of the scalar couplings, which we define at the electroweak scale ($\mt$). 
\begin{figure*}[t]
\includegraphics[width=0.49\linewidth]{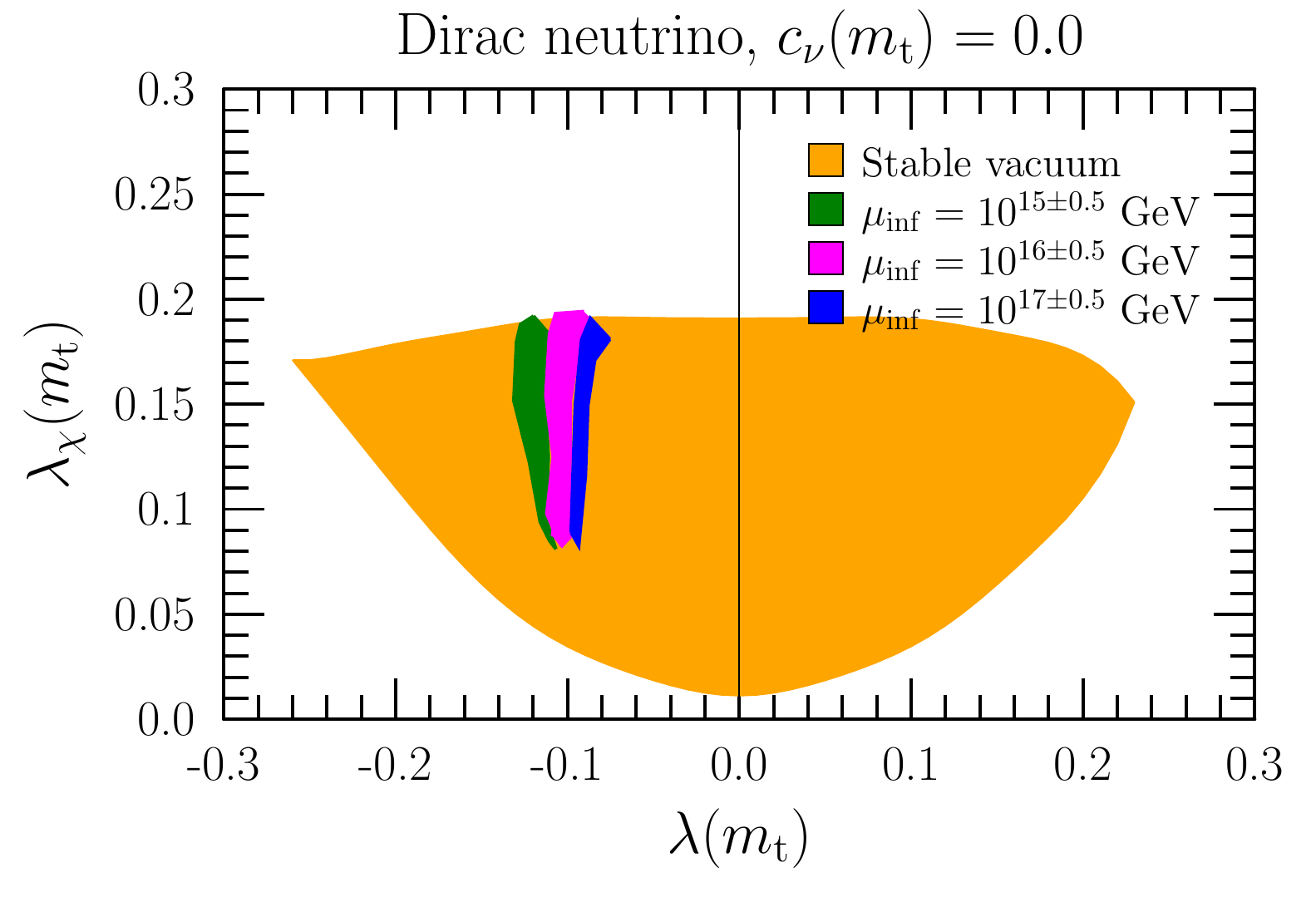}
\hfill
\includegraphics[width=0.49\linewidth]{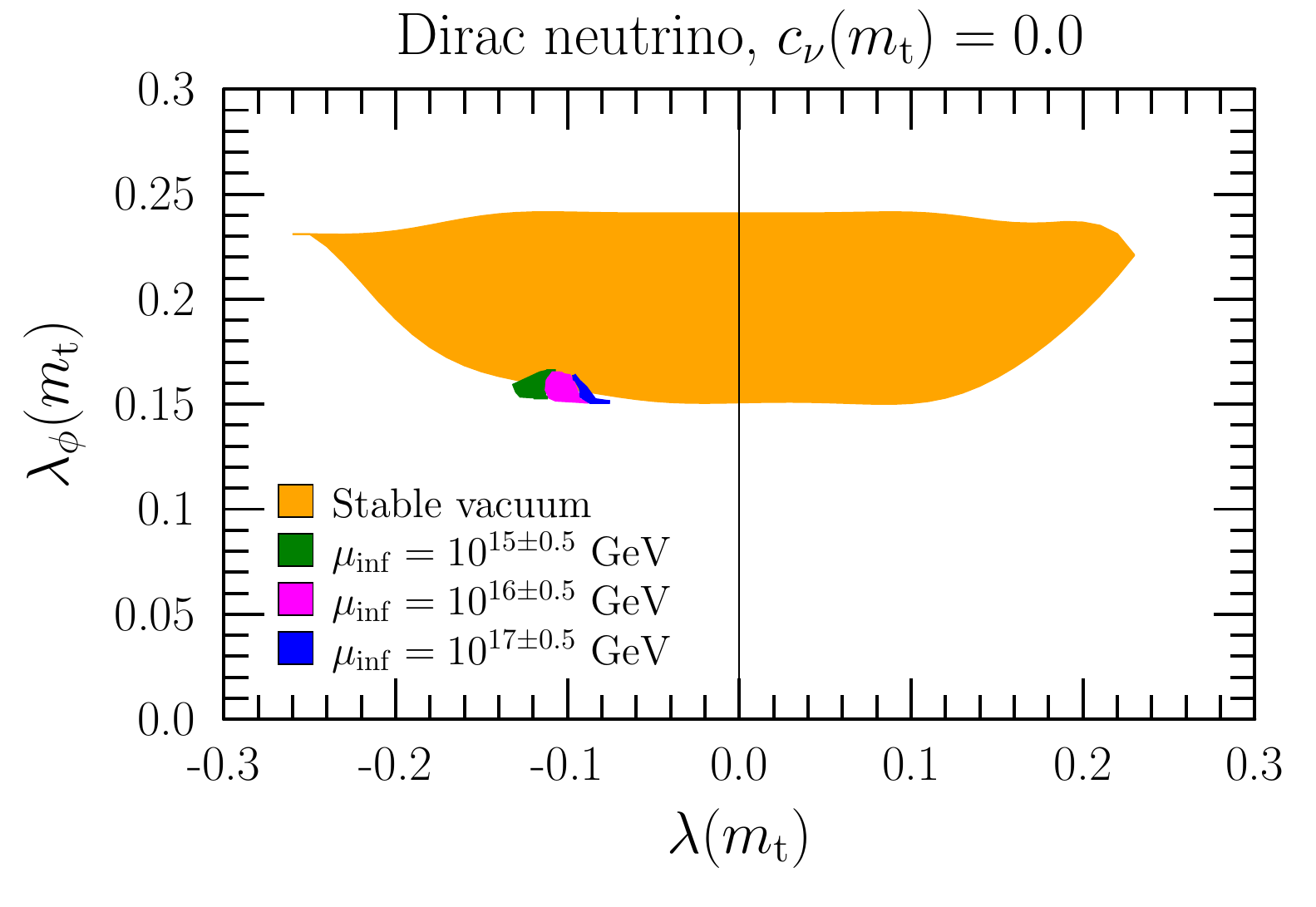}
\includegraphics[width=0.49\linewidth]{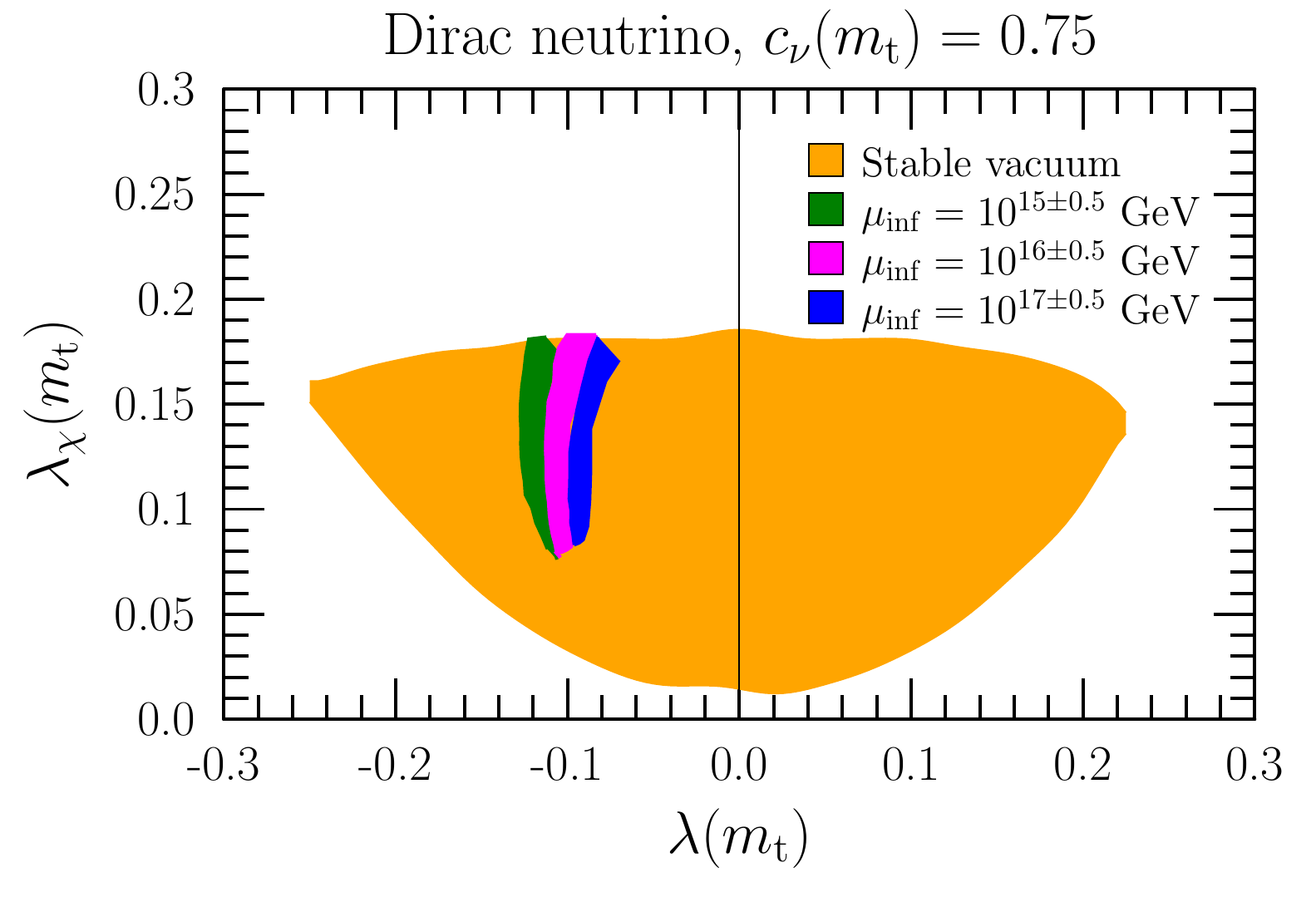}
\hfill
\includegraphics[width=0.49\linewidth]{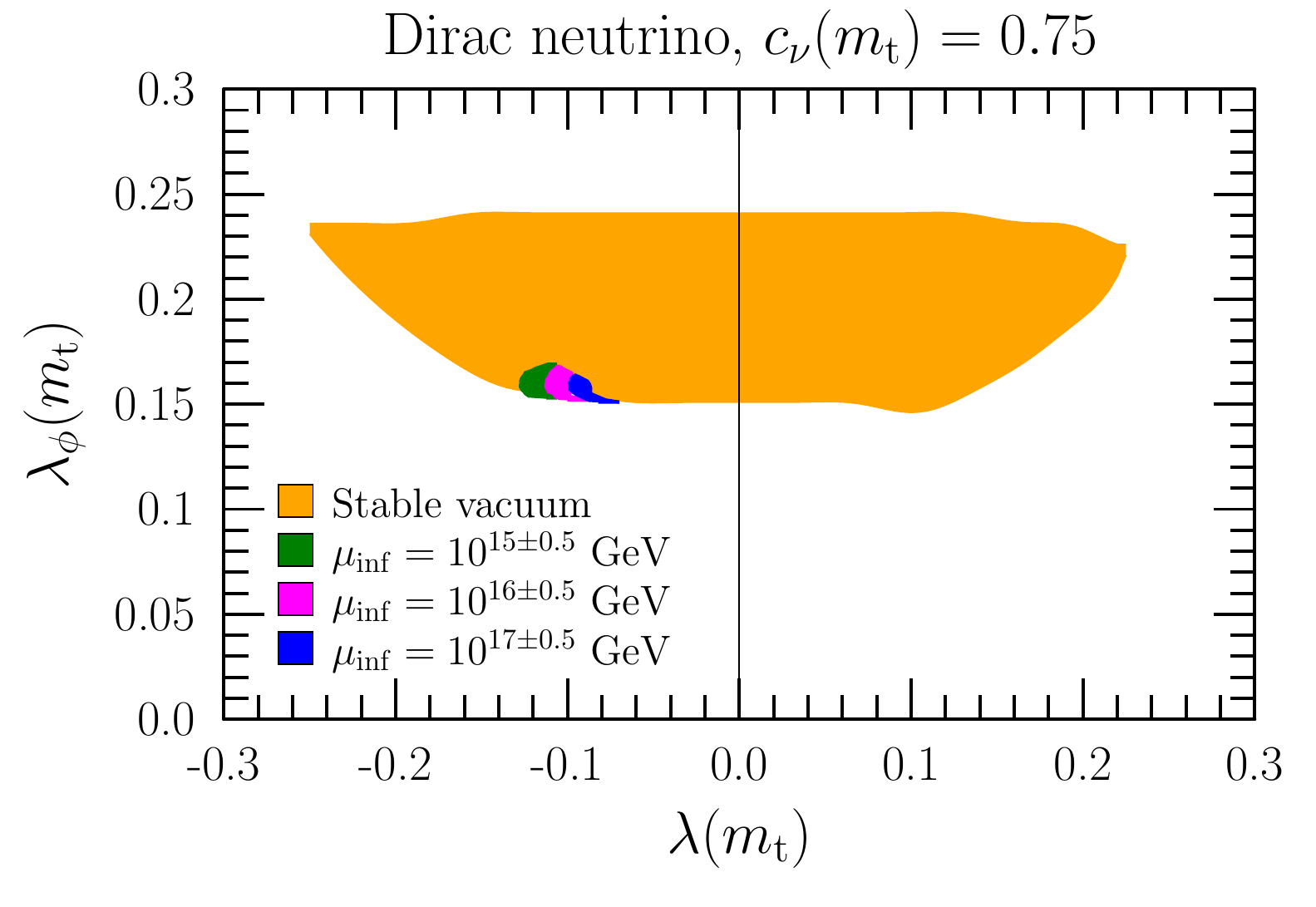}
\includegraphics[width=0.49\linewidth]{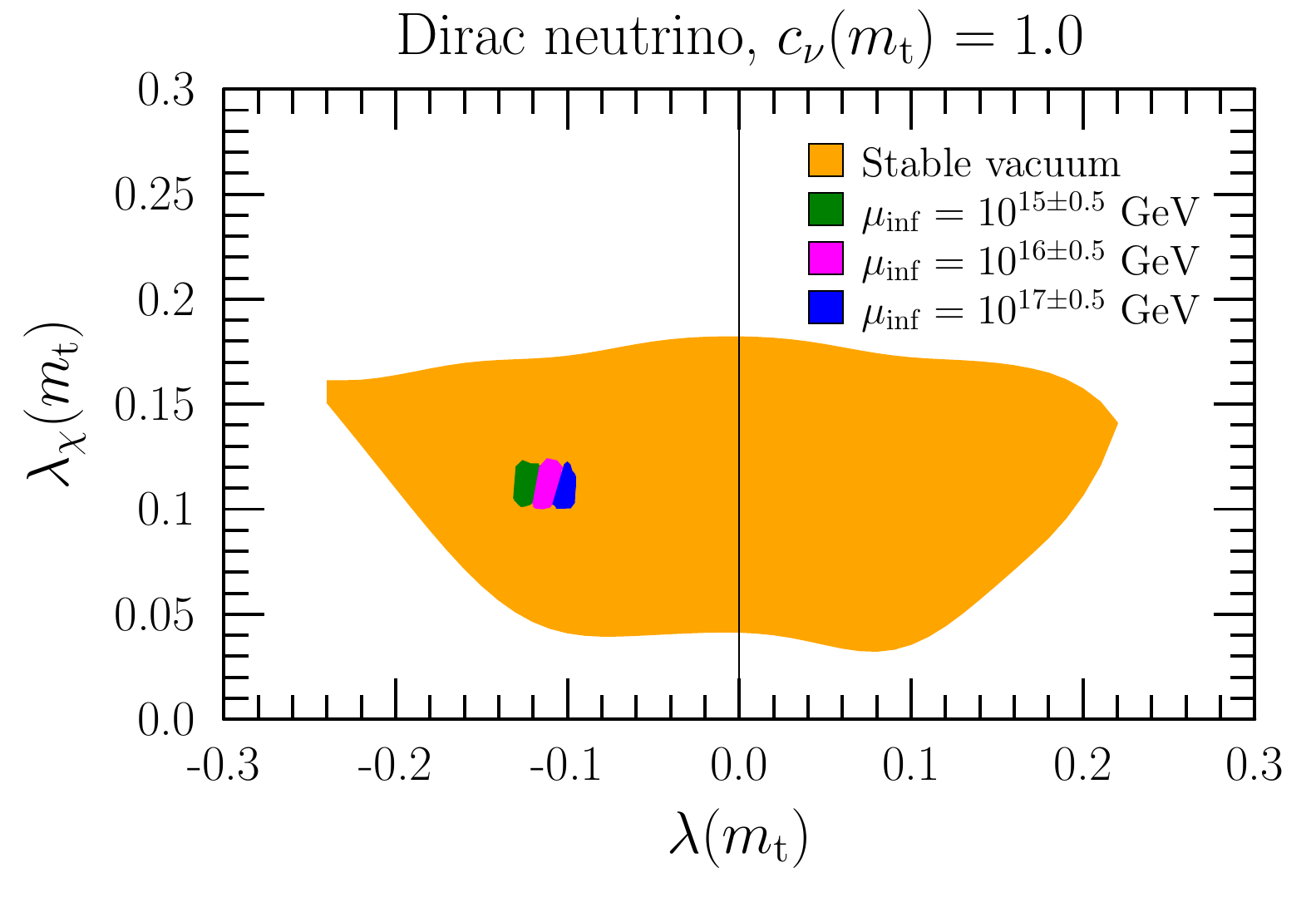}
\hfill
\includegraphics[width=.49\linewidth]{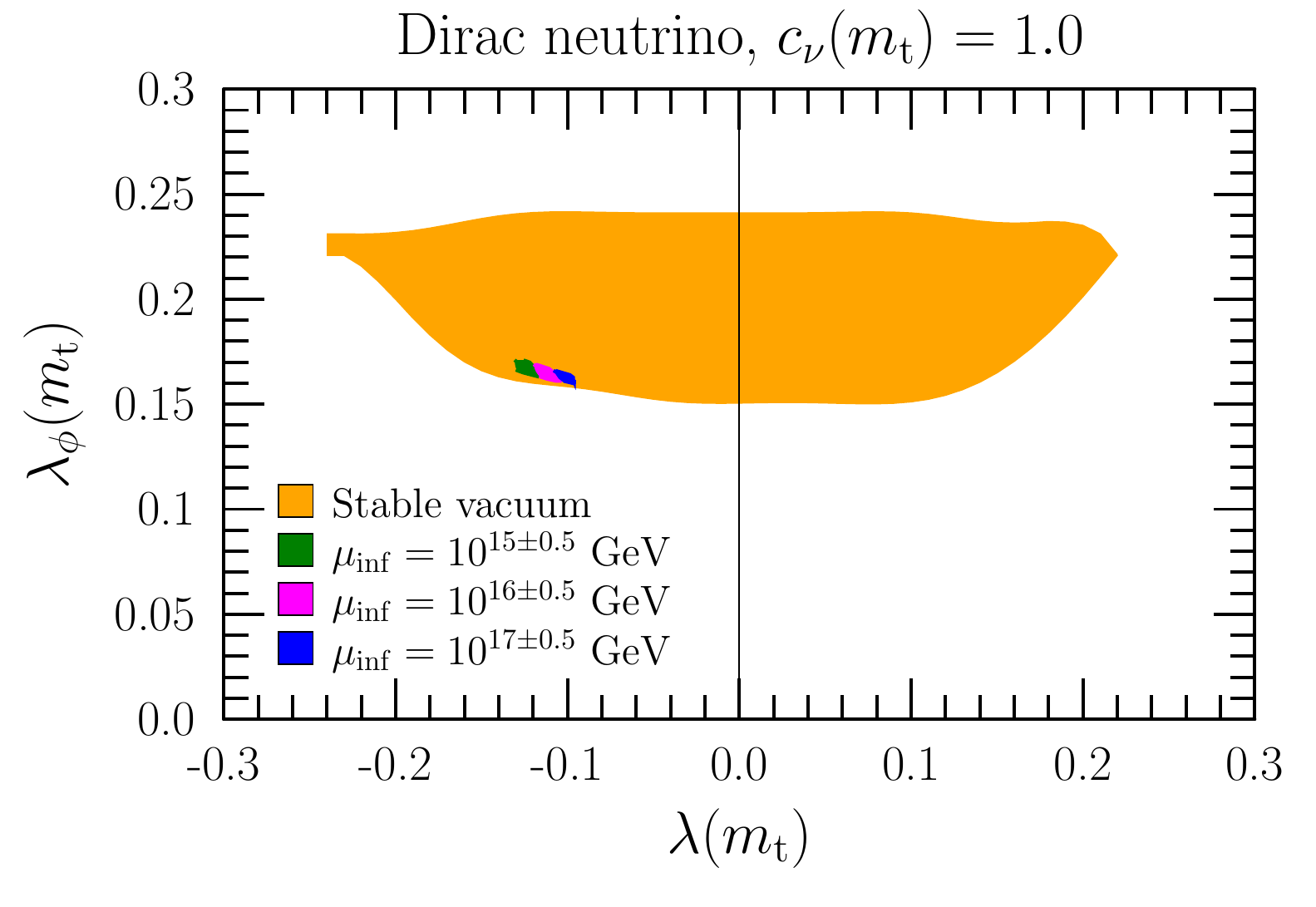}
\caption{\label{fig:projections}Projections of the allowed regions of the parameter space of scalar couplings onto the
$\lambda-\lambda_\chi$ plane (left) and $\lambda-\lambda_\phi$ plane (right) at three different ranges of the scale of inflation and at three selected values of the $c_\nu$ Yukawa coupling.}
\end{figure*}

The wedge in the running of $\detL$ appears only for $\lambda(m_\text{t})<0$. We have scanned this 
side of the parameter space by selecting fixed initial values for the scalar couplings 
$\lambda_\phi(m_\text{t})$, $\lambda_\chi(m_\text{t})$, and scanning the allowed initial values of the third one 
to find those points where the wedge in the running of $\detL$ appears with a minimum
$\lesssim\mathrm{O}(10^{-4})$. During this scanning, we need to search only for those points where $w(\mt)$ -- given by \cite{Peli:2019xwv}
$
w(\mt)  =
2 v(\mt) \sqrt{\frac{\displaystyle \lSM(\mt) (\lambda_\phi(\mt)-\lSM(\mt))}{\displaystyle (4 (\lambda_\phi(\mt) - \lSM(\mt)) \lambda_\chi(\mt) - \lambda(\mt)^2)}}
$,
with $\lSM(\mt) = \frac12 m_H(\mt)^2/v(\mt)^2 \simeq 0.126$ --
and the masses of the scalars given by \eqn{eq:MhH} remain positive.
We find that the parameter space is constrained to a shell on the surface of the region allowed by the 
conditions of  stability and perturbativity of $V$. The width of the shell is affected by the allowed depth of 
the minimum of $\detL$: the smaller $\detL$, the thinner the shell. Furthermore, 
we have also found that the minimum value of the location of the wedge $\muinf$ is around $10^{14}$ GeV, 
depending slightly on $c_\nu(m_\text{t})$.

In \fig{fig:projections} we present the results of such scan of the parameter space. These plots show different planar projections of the three dimensional parameter space, spanned by $\lambda_\phi(m_\text{t}),\lambda_\chi(m_\text{t})$ and $\lambda(m_\text{t})$. The shape and size of the supported regions is affected by the choice of $c_\nu(m_\text{t})$, as seen in the titles of the figures. We find that the parameter space of the scalar couplings is not empty, but constrained strongly if we assume that cosmic inflation took place as described above. This assumption constrains the smallest value of $w$ to around 265\,GeV.

%% file: PPMCISU.bbl
\providecommand{\noopsort}[1]{}\providecommand{\singleletter}[1]{#1}%
\begin{thebibliography}{24}%
\makeatletter
\providecommand \@ifxundefined [1]{%
 \@ifx{#1\undefined}
}%
\providecommand \@ifnum [1]{%
 \ifnum #1\expandafter \@firstoftwo
 \else \expandafter \@secondoftwo
 \fi
}%
\providecommand \@ifx [1]{%
 \ifx #1\expandafter \@firstoftwo
 \else \expandafter \@secondoftwo
 \fi
}%
\providecommand \natexlab [1]{#1}%
\providecommand \enquote  [1]{``#1''}%
\providecommand \bibnamefont  [1]{#1}%
\providecommand \bibfnamefont [1]{#1}%
\providecommand \citenamefont [1]{#1}%
\providecommand \href@noop [0]{\@secondoftwo}%
\providecommand \href [0]{\begingroup \@sanitize@url \@href}%
\providecommand \@href[1]{\@@startlink{#1}\@@href}%
\providecommand \@@href[1]{\endgroup#1\@@endlink}%
\providecommand \@sanitize@url [0]{\catcode `\\12\catcode `\$12\catcode
  `\&12\catcode `\#12\catcode `\^12\catcode `\_12\catcode `\%12\relax}%
\providecommand \@@startlink[1]{}%
\providecommand \@@endlink[0]{}%
\providecommand \url  [0]{\begingroup\@sanitize@url \@url }%
\providecommand \@url [1]{\endgroup\@href {#1}{\urlprefix }}%
\providecommand \urlprefix  [0]{URL }%
\providecommand \Eprint [0]{\href }%
\providecommand \doibase [0]{https://doi.org/}%
\providecommand \selectlanguage [0]{\@gobble}%
\providecommand \bibinfo  [0]{\@secondoftwo}%
\providecommand \bibfield  [0]{\@secondoftwo}%
\providecommand \translation [1]{[#1]}%
\providecommand \BibitemOpen [0]{}%
\providecommand \bibitemStop [0]{}%
\providecommand \bibitemNoStop [0]{.\EOS\space}%
\providecommand \EOS [0]{\spacefactor3000\relax}%
\providecommand \BibitemShut  [1]{\csname bibitem#1\endcsname}%
\let\auto@bib@innerbib\@empty
\bibitem [{\citenamefont {Guth}(1981)}]{Guth:1980zm}%
  \BibitemOpen
  \bibfield  {author} {\bibinfo {author} {\bibfnamefont {A.~H.}\ \bibnamefont
  {Guth}},\ }\bibfield  {title} {\bibinfo {title} {{The Inflationary Universe:
  A Possible Solution to the Horizon and Flatness Problems}},\ }\href
  {https://doi.org/10.1103/PhysRevD.23.347} {\bibfield  {journal} {\bibinfo
  {journal} {Phys. Rev.}\ }\textbf {\bibinfo {volume} {D23}},\ \bibinfo {pages}
  {347} (\bibinfo {year} {1981})},\ \bibinfo {note} {[Adv. Ser. Astrophys.
  Cosmol.3,139(1987)]}\BibitemShut {NoStop}%
\bibitem [{\citenamefont {Linde}(1982)}]{Linde:1981mu}%
  \BibitemOpen
  \bibfield  {author} {\bibinfo {author} {\bibfnamefont {A.~D.}\ \bibnamefont
  {Linde}},\ }\bibfield  {title} {\bibinfo {title} {{A New Inflationary
  Universe Scenario: A Possible Solution of the Horizon, Flatness, Homogeneity,
  Isotropy and Primordial Monopole Problems}},\ }\bibfield  {booktitle} {\emph
  {\bibinfo {booktitle} {{QUANTUM COSMOLOGY}}},\ }\href
  {https://doi.org/10.1016/0370-2693(82)91219-9} {\bibfield  {journal}
  {\bibinfo  {journal} {Phys. Lett.}\ }\textbf {\bibinfo {volume} {108B}},\
  \bibinfo {pages} {389} (\bibinfo {year} {1982})},\ \bibinfo {note} {[Adv.
  Ser. Astrophys. Cosmol.3,149(1987)]}\BibitemShut {NoStop}%
\bibitem [{\citenamefont {Albrecht}\ and\ \citenamefont
  {Steinhardt}(1982)}]{Albrecht:1982wi}%
  \BibitemOpen
  \bibfield  {author} {\bibinfo {author} {\bibfnamefont {A.}~\bibnamefont
  {Albrecht}}\ and\ \bibinfo {author} {\bibfnamefont {P.~J.}\ \bibnamefont
  {Steinhardt}},\ }\bibfield  {title} {\bibinfo {title} {{Cosmology for Grand
  Unified Theories with Radiatively Induced Symmetry Breaking}},\ }\href
  {https://doi.org/10.1103/PhysRevLett.48.1220} {\bibfield  {journal} {\bibinfo
   {journal} {Phys. Rev. Lett.}\ }\textbf {\bibinfo {volume} {48}},\ \bibinfo
  {pages} {1220} (\bibinfo {year} {1982})},\ \bibinfo {note} {[Adv. Ser.
  Astrophys. Cosmol.3,158(1987)]}\BibitemShut {NoStop}%
\bibitem [{\citenamefont {Baumann}(2011)}]{Baumann:2009ds}%
  \BibitemOpen
  \bibfield  {author} {\bibinfo {author} {\bibfnamefont {D.}~\bibnamefont
  {Baumann}},\ }\bibfield  {title} {\bibinfo {title} {{Inflation}},\ }in\ \href
  {https://doi.org/10.1142/9789814327183_0010} {\emph {\bibinfo {booktitle}
  {{Physics of the large and the small, TASI 09, proceedings of the Theoretical
  Advanced Study Institute in Elementary Particle Physics, Boulder, Colorado,
  USA, 1-26 June 2009}}}}\ (\bibinfo {year} {2011})\ pp.\ \bibinfo {pages}
  {523--686},\ \Eprint {https://arxiv.org/abs/0907.5424} {arXiv:0907.5424
  [hep-th]} \BibitemShut {NoStop}%
\bibitem [{\citenamefont {Linde}(1986)}]{Linde:1986fd}%
  \BibitemOpen
  \bibfield  {author} {\bibinfo {author} {\bibfnamefont {A.~D.}\ \bibnamefont
  {Linde}},\ }\bibfield  {title} {\bibinfo {title} {{Eternally Existing
  Selfreproducing Chaotic Inflationary Universe}},\ }\href
  {https://doi.org/10.1016/0370-2693(86)90611-8} {\bibfield  {journal}
  {\bibinfo  {journal} {Phys. Lett.}\ }\textbf {\bibinfo {volume} {B175}},\
  \bibinfo {pages} {395} (\bibinfo {year} {1986})}\BibitemShut {NoStop}%
\bibitem [{\citenamefont {Earman}\ and\ \citenamefont
  {Mosterin}(1999)}]{Earman:1999zia}%
  \BibitemOpen
  \bibfield  {author} {\bibinfo {author} {\bibfnamefont {J.}~\bibnamefont
  {Earman}}\ and\ \bibinfo {author} {\bibfnamefont {J.}~\bibnamefont
  {Mosterin}},\ }\bibfield  {title} {\bibinfo {title} {{A critical look at
  inflationary cosmology}},\ }\href@noop {} {\bibfield  {journal} {\bibinfo
  {journal} {Phil. Sci.}\ }\textbf {\bibinfo {volume} {66}},\ \bibinfo {pages}
  {1} (\bibinfo {year} {1999})}\BibitemShut {NoStop}%
\bibitem [{\citenamefont {Steinhardt}(2011)}]{Steinhardt:2011zza}%
  \BibitemOpen
  \bibfield  {author} {\bibinfo {author} {\bibfnamefont {P.~J.}\ \bibnamefont
  {Steinhardt}},\ }\bibfield  {title} {\bibinfo {title} {{The inflation debate:
  Is the theory at the heart of modern cosmology deeply flawed?}},\ }\href@noop
  {} {\bibfield  {journal} {\bibinfo  {journal} {Sci. Am.}\ }\textbf {\bibinfo
  {volume} {304N4}},\ \bibinfo {pages} {18} (\bibinfo {year}
  {2011})}\BibitemShut {NoStop}%
\bibitem [{\citenamefont {Aad}\ \emph {et~al.}(2012)\citenamefont {Aad} \emph
  {et~al.}}]{Aad:2012tfa}%
  \BibitemOpen
  \bibfield  {author} {\bibinfo {author} {\bibfnamefont {G.}~\bibnamefont
  {Aad}} \emph {et~al.} (\bibinfo {collaboration} {ATLAS}),\ }\bibfield
  {title} {\bibinfo {title} {{Observation of a new particle in the search for
  the Standard Model Higgs boson with the ATLAS detector at the LHC}},\ }\href
  {https://doi.org/10.1016/j.physletb.2012.08.020} {\bibfield  {journal}
  {\bibinfo  {journal} {Phys. Lett.}\ }\textbf {\bibinfo {volume} {B716}},\
  \bibinfo {pages} {1} (\bibinfo {year} {2012})},\ \Eprint
  {https://arxiv.org/abs/1207.7214} {arXiv:1207.7214 [hep-ex]} \BibitemShut
  {NoStop}%
\bibitem [{\citenamefont {Chatrchyan}\ \emph {et~al.}(2012)\citenamefont
  {Chatrchyan} \emph {et~al.}}]{Chatrchyan:2012xdj}%
  \BibitemOpen
  \bibfield  {author} {\bibinfo {author} {\bibfnamefont {S.}~\bibnamefont
  {Chatrchyan}} \emph {et~al.} (\bibinfo {collaboration} {CMS}),\ }\bibfield
  {title} {\bibinfo {title} {{Observation of a New Boson at a Mass of 125 GeV
  with the CMS Experiment at the LHC}},\ }\href
  {https://doi.org/10.1016/j.physletb.2012.08.021} {\bibfield  {journal}
  {\bibinfo  {journal} {Phys. Lett.}\ }\textbf {\bibinfo {volume} {B716}},\
  \bibinfo {pages} {30} (\bibinfo {year} {2012})},\ \Eprint
  {https://arxiv.org/abs/1207.7235} {arXiv:1207.7235 [hep-ex]} \BibitemShut
  {NoStop}%
\bibitem [{\citenamefont {Bezrukov}\ \emph {et~al.}(2015)\citenamefont
  {Bezrukov}, \citenamefont {Rubio},\ and\ \citenamefont
  {Shaposhnikov}}]{Bezrukov:2014ipa}%
  \BibitemOpen
  \bibfield  {author} {\bibinfo {author} {\bibfnamefont {F.}~\bibnamefont
  {Bezrukov}}, \bibinfo {author} {\bibfnamefont {J.}~\bibnamefont {Rubio}},\
  and\ \bibinfo {author} {\bibfnamefont {M.}~\bibnamefont {Shaposhnikov}},\
  }\bibfield  {title} {\bibinfo {title} {{Living beyond the edge: Higgs
  inflation and vacuum metastability}},\ }\href
  {https://doi.org/10.1103/PhysRevD.92.083512} {\bibfield  {journal} {\bibinfo
  {journal} {Phys. Rev.}\ }\textbf {\bibinfo {volume} {D92}},\ \bibinfo {pages}
  {083512} (\bibinfo {year} {2015})},\ \Eprint
  {https://arxiv.org/abs/1412.3811} {arXiv:1412.3811 [hep-ph]} \BibitemShut
  {NoStop}%
\bibitem [{\citenamefont {Martin}\ \emph {et~al.}(2014)\citenamefont {Martin},
  \citenamefont {Ringeval},\ and\ \citenamefont {Vennin}}]{Martin:2013tda}%
  \BibitemOpen
  \bibfield  {author} {\bibinfo {author} {\bibfnamefont {J.}~\bibnamefont
  {Martin}}, \bibinfo {author} {\bibfnamefont {C.}~\bibnamefont {Ringeval}},\
  and\ \bibinfo {author} {\bibfnamefont {V.}~\bibnamefont {Vennin}},\
  }\bibfield  {title} {\bibinfo {title} {{Encyclopædia Inflationaris}},\
  }\href {https://doi.org/10.1016/j.dark.2014.01.003} {\bibfield  {journal}
  {\bibinfo  {journal} {Phys. Dark Univ.}\ }\textbf {\bibinfo {volume} {5-6}},\
  \bibinfo {pages} {75} (\bibinfo {year} {2014})},\ \Eprint
  {https://arxiv.org/abs/1303.3787} {arXiv:1303.3787 [astro-ph.CO]}
  \BibitemShut {NoStop}%
\bibitem [{\citenamefont {Ade}\ \emph {et~al.}(2016{\natexlab{a}})\citenamefont
  {Ade} \emph {et~al.}}]{Ade:2015lrj}%
  \BibitemOpen
  \bibfield  {author} {\bibinfo {author} {\bibfnamefont {P.~A.~R.}\
  \bibnamefont {Ade}} \emph {et~al.} (\bibinfo {collaboration} {Planck}),\
  }\bibfield  {title} {\bibinfo {title} {{Planck 2015 results. XX. Constraints
  on inflation}},\ }\href {https://doi.org/10.1051/0004-6361/201525898}
  {\bibfield  {journal} {\bibinfo  {journal} {Astron. Astrophys.}\ }\textbf
  {\bibinfo {volume} {594}},\ \bibinfo {pages} {A20} (\bibinfo {year}
  {2016}{\natexlab{a}})},\ \Eprint {https://arxiv.org/abs/1502.02114}
  {arXiv:1502.02114 [astro-ph.CO]} \BibitemShut {NoStop}%
\bibitem [{\citenamefont {Ade}\ \emph {et~al.}(2016{\natexlab{b}})\citenamefont
  {Ade} \emph {et~al.}}]{Array:2015xqh}%
  \BibitemOpen
  \bibfield  {author} {\bibinfo {author} {\bibfnamefont {P.~A.~R.}\
  \bibnamefont {Ade}} \emph {et~al.} (\bibinfo {collaboration} {BICEP2, Keck
  Array}),\ }\bibfield  {title} {\bibinfo {title} {{Improved Constraints on
  Cosmology and Foregrounds from BICEP2 and Keck Array Cosmic Microwave
  Background Data with Inclusion of 95 GHz Band}},\ }\href
  {https://doi.org/10.1103/PhysRevLett.116.031302} {\bibfield  {journal}
  {\bibinfo  {journal} {Phys. Rev. Lett.}\ }\textbf {\bibinfo {volume} {116}},\
  \bibinfo {pages} {031302} (\bibinfo {year} {2016}{\natexlab{b}})},\ \Eprint
  {https://arxiv.org/abs/1510.09217} {arXiv:1510.09217 [astro-ph.CO]}
  \BibitemShut {NoStop}%
\bibitem [{\citenamefont {Mollerach}(1990)}]{Mollerach:1989hu}%
  \BibitemOpen
  \bibfield  {author} {\bibinfo {author} {\bibfnamefont {S.}~\bibnamefont
  {Mollerach}},\ }\bibfield  {title} {\bibinfo {title} {{Isocurvature Baryon
  Perturbations and Inflation}},\ }\href
  {https://doi.org/10.1103/PhysRevD.42.313} {\bibfield  {journal} {\bibinfo
  {journal} {Phys. Rev.}\ }\textbf {\bibinfo {volume} {D42}},\ \bibinfo {pages}
  {313} (\bibinfo {year} {1990})}\BibitemShut {NoStop}%
\bibitem [{\citenamefont {Linde}\ and\ \citenamefont
  {Mukhanov}(1997)}]{Linde:1996gt}%
  \BibitemOpen
  \bibfield  {author} {\bibinfo {author} {\bibfnamefont {A.~D.}\ \bibnamefont
  {Linde}}\ and\ \bibinfo {author} {\bibfnamefont {V.~F.}\ \bibnamefont
  {Mukhanov}},\ }\bibfield  {title} {\bibinfo {title} {{Nongaussian
  isocurvature perturbations from inflation}},\ }\href
  {https://doi.org/10.1103/PhysRevD.56.R535} {\bibfield  {journal} {\bibinfo
  {journal} {Phys. Rev.}\ }\textbf {\bibinfo {volume} {D56}},\ \bibinfo {pages}
  {R535} (\bibinfo {year} {1997})},\ \Eprint
  {https://arxiv.org/abs/astro-ph/9610219} {arXiv:astro-ph/9610219 [astro-ph]}
  \BibitemShut {NoStop}%
\bibitem [{\citenamefont {Lyth}\ and\ \citenamefont
  {Wands}(2002)}]{Lyth:2001nq}%
  \BibitemOpen
  \bibfield  {author} {\bibinfo {author} {\bibfnamefont {D.~H.}\ \bibnamefont
  {Lyth}}\ and\ \bibinfo {author} {\bibfnamefont {D.}~\bibnamefont {Wands}},\
  }\bibfield  {title} {\bibinfo {title} {{Generating the curvature perturbation
  without an inflaton}},\ }\href
  {https://doi.org/10.1016/S0370-2693(01)01366-1} {\bibfield  {journal}
  {\bibinfo  {journal} {Phys. Lett.}\ }\textbf {\bibinfo {volume} {B524}},\
  \bibinfo {pages} {5} (\bibinfo {year} {2002})},\ \Eprint
  {https://arxiv.org/abs/hep-ph/0110002} {arXiv:hep-ph/0110002 [hep-ph]}
  \BibitemShut {NoStop}%
\bibitem [{\citenamefont {Kitajima}\ \emph {et~al.}(2017)\citenamefont
  {Kitajima}, \citenamefont {Langlois}, \citenamefont {Takahashi},\ and\
  \citenamefont {Yokoyama}}]{Kitajima:2017fiy}%
  \BibitemOpen
  \bibfield  {author} {\bibinfo {author} {\bibfnamefont {N.}~\bibnamefont
  {Kitajima}}, \bibinfo {author} {\bibfnamefont {D.}~\bibnamefont {Langlois}},
  \bibinfo {author} {\bibfnamefont {T.}~\bibnamefont {Takahashi}},\ and\
  \bibinfo {author} {\bibfnamefont {S.}~\bibnamefont {Yokoyama}},\ }\bibfield
  {title} {\bibinfo {title} {{Refined Study of Isocurvature Fluctuations in the
  Curvaton Scenario}},\ }\href {https://doi.org/10.1088/1475-7516/2017/12/042}
  {\bibfield  {journal} {\bibinfo  {journal} {JCAP}\ }\textbf {\bibinfo
  {volume} {1712}}\bibfield  {number} {\bibinfo  {number} { (12)},\ \bibinfo
  {pages} {042}},\ }\Eprint {https://arxiv.org/abs/1707.06929}
  {arXiv:1707.06929 [astro-ph.CO]} \BibitemShut {NoStop}%
\bibitem [{Note1()}]{Note1}%
  \BibitemOpen
  \bibinfo {note} {For $\lambda >0$ the global minimum of $V$ falls on one of
  the coordinate axes when $\protect \ensuremath {{\protect \qopname \relax
  m{det}}_{C}}$ becomes small. As a result our inflationary scenario does not
  work.}\BibitemShut {Stop}%
\bibitem [{\citenamefont {P\'eli}\ and\ \citenamefont
  {Tr\'ocs\'anyi}(2019)}]{Peli:2019xwv}%
  \BibitemOpen
  \bibfield  {author} {\bibinfo {author} {\bibfnamefont {Z.}~\bibnamefont
  {P\'eli}}\ and\ \bibinfo {author} {\bibfnamefont {Z.}~\bibnamefont
  {Tr\'ocs\'anyi}},\ }\bibfield  {title} {\bibinfo {title} {{Stability of the
  vacuum as constraint on $U$(1) extensions of the standard model}},\
  }\href@noop {} {\  (\bibinfo {year} {2019})},\ \Eprint
  {https://arxiv.org/abs/1902.02791} {arXiv:1902.02791 [hep-ph]} \BibitemShut
  {NoStop}%
\bibitem [{Note2()}]{Note2}%
  \BibitemOpen
  \bibinfo {note} {In our nodel for inflation we only consider the case when
  both $\mu ^2_i>0$.}\BibitemShut {Stop}%
\bibitem [{Note3()}]{Note3}%
  \BibitemOpen
  \bibinfo {note} {The would be massless Goldstone boson can give mass to a new
  neutral vector boson as described in Ref.\protect \tmspace +\thinmuskip
  {.1667em}\cite {Trocsanyi:2018bkm}}\BibitemShut {NoStop}%
\bibitem [{\citenamefont {Byrnes}\ and\ \citenamefont
  {Wands}(2006)}]{Byrnes:2006fr}%
  \BibitemOpen
  \bibfield  {author} {\bibinfo {author} {\bibfnamefont {C.~T.}\ \bibnamefont
  {Byrnes}}\ and\ \bibinfo {author} {\bibfnamefont {D.}~\bibnamefont {Wands}},\
  }\bibfield  {title} {\bibinfo {title} {{Curvature and isocurvature
  perturbations from two-field inflation in a slow-roll expansion}},\ }\href
  {https://doi.org/10.1103/PhysRevD.74.043529} {\bibfield  {journal} {\bibinfo
  {journal} {Phys. Rev.}\ }\textbf {\bibinfo {volume} {D74}},\ \bibinfo {pages}
  {043529} (\bibinfo {year} {2006})},\ \Eprint
  {https://arxiv.org/abs/astro-ph/0605679} {arXiv:astro-ph/0605679 [astro-ph]}
  \BibitemShut {NoStop}%
\bibitem [{\citenamefont {Gordon}\ \emph {et~al.}(2001)\citenamefont {Gordon},
  \citenamefont {Wands}, \citenamefont {Bassett},\ and\ \citenamefont
  {Maartens}}]{Gordon:2000hv}%
  \BibitemOpen
  \bibfield  {author} {\bibinfo {author} {\bibfnamefont {C.}~\bibnamefont
  {Gordon}}, \bibinfo {author} {\bibfnamefont {D.}~\bibnamefont {Wands}},
  \bibinfo {author} {\bibfnamefont {B.~A.}\ \bibnamefont {Bassett}},\ and\
  \bibinfo {author} {\bibfnamefont {R.}~\bibnamefont {Maartens}},\ }\bibfield
  {title} {\bibinfo {title} {{Adiabatic and entropy perturbations from
  inflation}},\ }\href {https://doi.org/10.1103/PhysRevD.63.023506} {\bibfield
  {journal} {\bibinfo  {journal} {Phys. Rev.}\ }\textbf {\bibinfo {volume}
  {D63}},\ \bibinfo {pages} {023506} (\bibinfo {year} {2001})},\ \Eprint
  {https://arxiv.org/abs/astro-ph/0009131} {arXiv:astro-ph/0009131 [astro-ph]}
  \BibitemShut {NoStop}%
\bibitem [{\citenamefont {Tr\'ocs\'anyi}(2018)}]{Trocsanyi:2018bkm}%
  \BibitemOpen
  \bibfield  {author} {\bibinfo {author} {\bibfnamefont {Z.}~\bibnamefont
  {Tr\'ocs\'anyi}},\ }\bibfield  {title} {\bibinfo {title} {{Super-weak force
  and neutrino masses}},\ }\href@noop {} {\  (\bibinfo {year} {2018})},\
  \Eprint {https://arxiv.org/abs/1812.11189} {arXiv:1812.11189 [hep-ph]}
  \BibitemShut {NoStop}%
\end{thebibliography}%
